# Highly Multiplexible Thermal Kinetic Inductance Detectors for X-Ray Imaging Spectroscopy

Gerhard Ulbricht[1,a)], Benjamin A. Mazin[1], Paul Szypryt[1], Alex B. Walter[1], Clint Bockstiegel[1], Bruce Bumble[2]

[1] *Department of Physics, University of California, Santa Barbara, California 93106, USA*
[2] *NASA Jet Propulsion Laboratory, 4800 Oak Grove Drive, Pasadena, California 91125, USA*

For X-ray imaging spectroscopy, high spatial resolution over a large field of view is often as important as high energy resolution, but current X-ray detectors do not provide both in the same device. Thermal Kinetic Inductance Detectors (TKIDs) are being developed as they offer a feasible way to combine the energy resolution of transition edge sensors with pixel counts approaching CCDs and thus promise significant improvements for many X-ray spectroscopy applications. TKIDs are a variation of Microwave Kinetic Inductance Detectors (MKIDs) and share their multiplexibility: working MKID arrays with 2024 pixels have recently been demonstrated and much bigger arrays are under development. In this work, we present our first working TKID prototypes which are able to achieve an energy resolution of 75 eV at 5.9 keV, even though their general design still has to be optimized. We further describe TKID fabrication, characterization, multiplexing and working principle and demonstrate the necessity of a data fitting algorithm in order to extract photon energies. With further design optimizations we expect to be able to improve our TKID energy resolution to less than 10 eV at 5.9 keV.

Thermal Kinetic Inductance Detectors (TKIDs) are a new type of X-ray detector[1,2]. They have the potential to achieve excellent time and energy resolution and, most importantly, can be scaled up to kilo- or even mega-pixel arrays[3]. TKIDs promise enormous advancement in X-ray imaging spectroscopy, a powerful and versatile tool to study scientific questions as diverse as the astrophysical processes around black holes[4], understanding of semiconductor fabrication irregularities[5], and elemental identification in biological samples[6]. Many modern applications of X-ray imaging spectroscopy require both high energy and spatial resolution over a large field of view. Excellent energy resolution (down to 0.5 eV at 5.9 keV[7]) can be achieved with diffractometers, but they have to be scanned mechanically over their energy range creating an impractical time penalty, and they are not applicable to spatially extended X-ray sources such as supernova remnants. For these reasons, Transition Edge Sensor (TES) arrays are currently the dominant spatially resolving X-ray detectors. They still resolve an impressive 2.4 eV at 5.9 keV[8], and arrays with more than 1000 pixels are under development. However, scaling TESs up to larger pixel counts is very challenging. As they operate below 100 mK, it is impossible to connect several thousand TESs with 3 wires each to room temperature, forcing larger TES arrays to use complex multiplexing schemes. Impressively elaborate approaches[9,10] have been demonstrated for TES multiplexing, but the sheer complexity necessary makes more than several kilopixels impracticable. TKIDs offer a straightforward and feasible route to kilo- or even mega-pixel arrays. Further development could improve TKID energy resolution to rival TESs, while future TES multiplexing is unlikely to scale to the pixel counts TKIDs can reach.

TKIDs are a variation of Microwave Kinetic Inductance Detectors (MKIDs[11-14]). Every TKID pixel is formed by combining a superconducting capacitor and inductor into an LC resonator circuit with a unique, lithographically defined resonance frequency, coupled to a microwave feed line. More than 2000 resonators can be coupled to a single feed line and amplifier, allowing them to be read out simultaneously with just two microwave lines (signal in/signal out) connecting the sensors from around 100 mK to room temperature. Working MKID arrays with 2024 pixels have already been demonstrated[15], and significant further array growth is restrained more by the cost of room temperature readout electronics (about $ 10 per pixel[16] at the moment) than by fundamental scientific challenges.

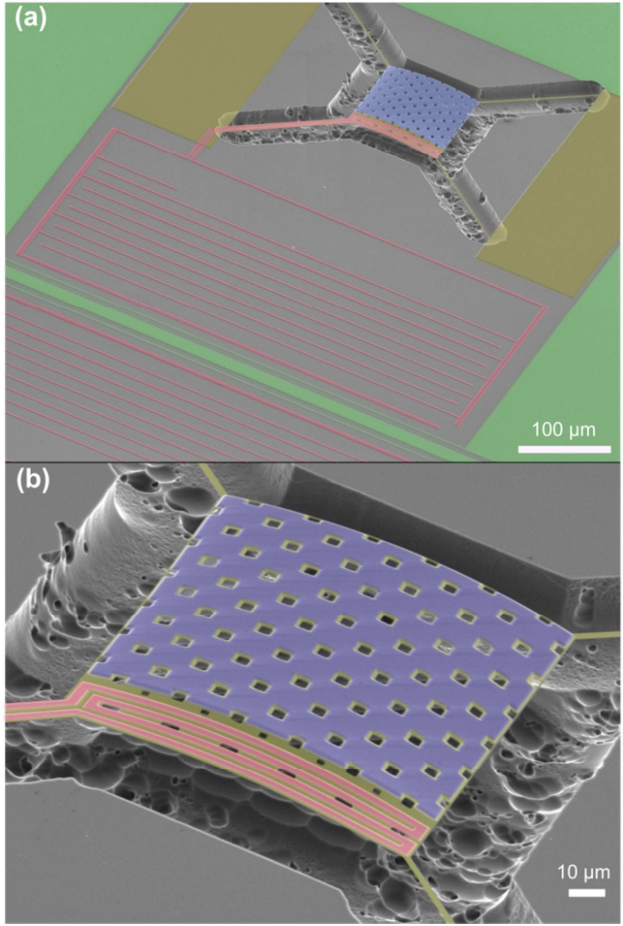

*Fig. 1: a.: A TKID on a silicon substrate. Interdigitated capacitors and meandered inductors are etched from 200nm of sub-stoichiometric $TiN_x$ (red – colors added for clarity). These form LC circuits, each with a unique resonance frequency, coupled to a CPW Nb (green) feed line. The inductors, and separate 500 nm thick Ta (blue) absorbers sit on a free-standing $Si_3N_4$ (yellow) island.*

*b.: SEM of a TKID island. The Ta absorber and the $Si_3N_4$ beneath it are perforated to facilitate under-etching with $XeF_2$. We etch about 10 – 15 µm deep into the Si. The thin $Si_3N_4$ bridges holding the free-standing island have a cross section of 2.0 µm x 0.25 µm and are 141 µm long.*

Our TKIDs are fabricated from a 200 nm thick sub-stoichiometric $TiN_x$ film[17], a Nb ground plane and feed line, and a 500 nm thick superconducting Ta absorber that will absorb about 26% of all passing Mn K-α X-ray photons[18]; see Fig. 1. A free-standing $Si_3N_4$ membrane is used to thermally isolate the absorber and a part of the TKID inductor (for further details please see supplementary materials[19]). We use sub-stoichiometric $TiN_x$ ($x < 1$)[20] in order to obtain a 1.6 K critical temperature ($T_c$). We operate our TKIDs between 75 mK and 200 mK, well below $T_c$, and expect the membrane temperature to rise by about 200-300 mK when a 6 keV photon is absorbed. In future development we intend to use mushroom shaped absorbers[21,22] to cover the entire resonator pixel, eliminating X-ray photon hits in undesired locations and increasing the pixel fill factor. However, to demonstrate the device we have started with a simpler design that still allows many photons to hit the Si substrate. These produce pulse shapes clearly different from hits on the Ta absorber, allowing us to filter out these substrate hits during data analysis.

TKID resonators are read out by driving them at their respective resonance frequencies and monitoring the phase between excitation and oscillation[14,23]. They generate a signal pulse for each individual photon absorbed (Fig. 2)[19] and operate as microcalorimeters[24]: when a photon hits the absorber, the temperature of the free-standing island increases. This temperature rise breaks Cooper pairs and creates unpaired single electron excitations (so called quasiparticles) in the superconducting $TiN_x$ inductor on the island. As Cooper pair density and kinetic inductance are coupled, the kinetic inductance and with it the resonator's total inductance increases. This causes its resonance frequency to decrease (Fig. 3.c) and, as the drive frequency stays constant, the phase between oscillation and excitation to shift (Fig. 3.a). We plot this phase shift over time and get a unique pulse for every absorbed photon (Fig. 2.b), allowing us to not only determine photon number and arrival times but also the energy of every single photon from the maximum island temperature it provoked (see below). As is conventional with MKIDs[23], we measure phase not with respect to the origin of the IQ plane, but to the center of the resonance loop (Fig. 2.a). Thus, the resonator phase shifts from π to –π during a frequency sweep.

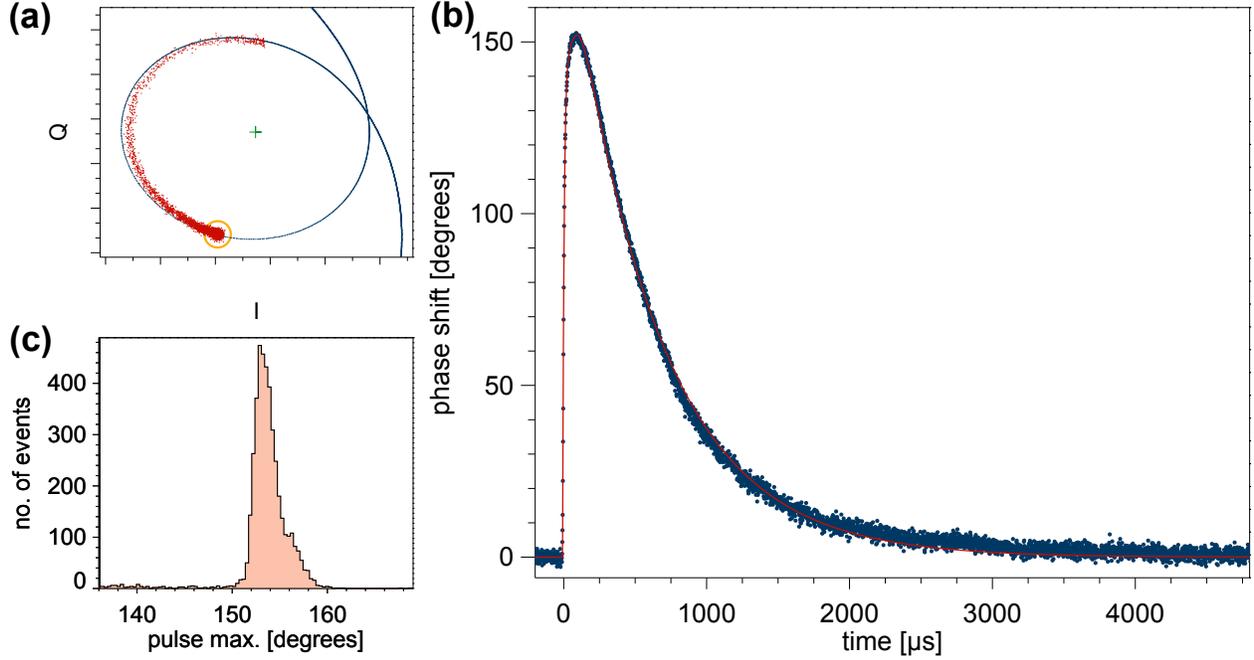

*Fig. 2: a.: Signal (red dots) caused by absorption of an X-ray photon, shown in the I/Q plane of the GHz drive voltage. Also shown are the I/Q values (blue line) of a frequency sweep extrapolated from a lower power measurement and the loop center (green cross). During the measurement, I/Q is constant (orange circle) until a photon hits, then it shifts quickly to the upper part of the loop and slowly relaxes afterwards.*

*b.: Phase pulse of a photon hit (blue dots) extracted from the I/Q plane. After photon absorption, it takes several microseconds for the TKID island to thermalize and the phase shift to reach a maximum. As the island is thermally isolated the subsequent pulse decay time is much longer than the quasiparticle recombination time and mainly given by the island's thermal coupling to the heat bath. The overplotted red line is the fit used to calculate the photon energy (see text). In addition to counting single photons and measuring their energies, TKIDs determine photon arrival times with microsecond accuracy.*

*c.: Distribution of the maximum pulse heights of 4970 X-ray photons (same data set as Fig. 4), demonstrating the necessity of pulse fitting. As the maximum pulse height is a poor determinant for photon energy, the entire pulse shape needs to be considered.*

A photon's energy is coded into the maximum of its phase pulse as that corresponds to the maximum island temperature reached. However, since every pulse has a slightly different rise time (likely caused by varying absorption locations) and thermal energy is lost to the bath even during thermalization, analyzing just the maximum pulse height is insufficient (Fig. 2.c). Instead, we fit every pulse in order to reconstruct the maximum signal at the moment of photon absorption. To better illustrate, Fig. 3.a shows four frequency sweeps of the same resonator at different temperatures. Photon absorption increases the island temperature and thus causes this frequency response curve to shift to one of the higher temperature values and then relax back. A phase pulse, as shown in Fig. 2.b, is generated by monitoring the shifting frequency response at a constant measurement frequency. A steeper frequency response curve thus results in higher resonator sensitivities (but also reduces the dynamic range). The slope of the response curve is dominated by the resonator's measured quality factor, $Q_m$, which consists of the internal quality factor, $Q_i$, and the coupling quality factor, $Q_c$ ($1/Q_m = 1/Q_i + 1/Q_c$). Therefore, we require a superconductor with high $Q_i$ to achieve high sensitivity[19].

Fig. 3.c shows the dependence of our TKIDs' resonance frequency, $f_r$, on temperature as well as a fit[23,24] (based on Mattis-Bardeen theory) for a $T_c$ of 1.6 K. In the temperature range our TKIDs operate in, this $f_r$ (T) behavior can be approximated as two linear regions with different slopes. Since after absorbing an X-ray photon the island cools exponentially, these two slopes cause the phase pulse to exhibit two different exponential decay times. Therefore we use a double exponential decay to fit TKID phase pulses. We found

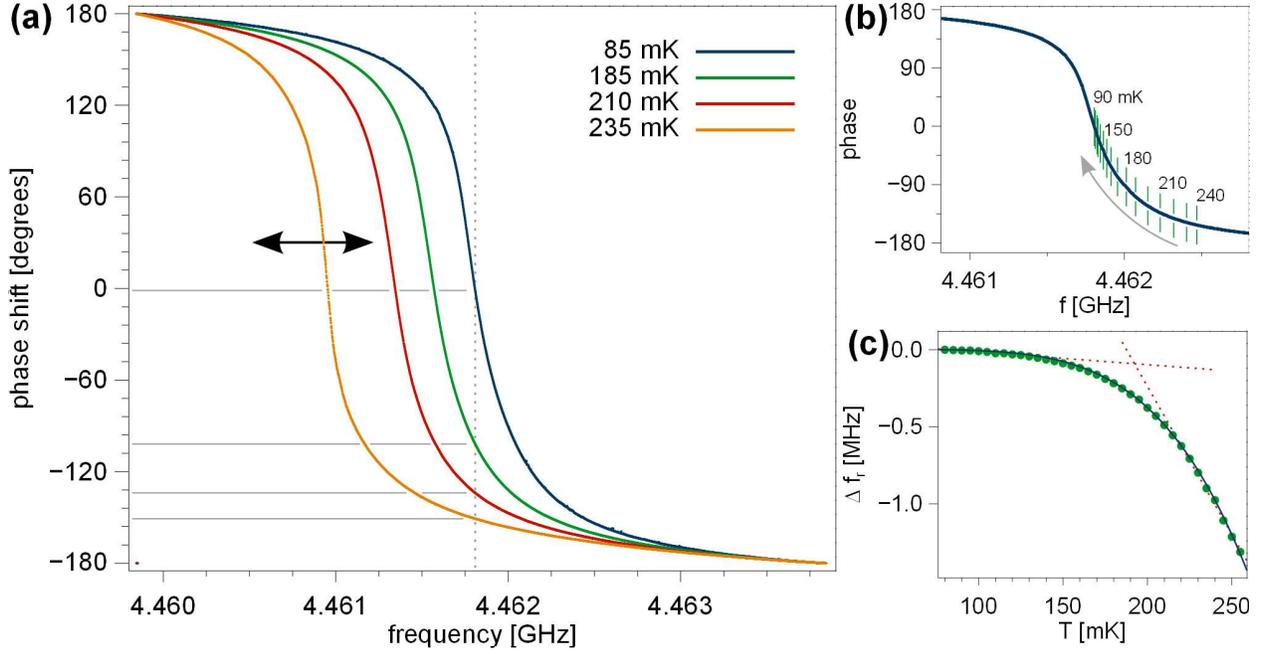

*Fig. 3: a.: Frequency sweeps at 85, 185, 210 and 235 mK. During a measurement, the resonator is driven with its resonance frequency at base temperature (dotted vertical line). Photon absorption shifts this frequency response curve to a higher temperature value (indicated by the arrow), followed by relaxation. As the response curve shape varies little with temperature, phase pulses like Fig. 2.b can be understood as moving along a frequency response curve and measuring the respective phase values (see 3.b).*

*b.: Green lines are 10 mK apart in temperature and represent the expected phase values (extracted from 3.a) along the 85 mK frequency response curve (blue line). They vary in distance as resonance frequency vs. temperature (3.c) is not linear. The arrow indicates the exponential cooling of the free-standing island.*

*c.: Decrease of resonance frequency $f_r$ with increasing temperature T (green circles) and expected dependence for a $T_c$ of 1.6 K (blue line). In the temperature region our TKIDs operate in, $f_r(T)$ can be approximated as 2 linear regimes with different slopes (indicated by dotted red lines), causing (together with the distinct shape of the frequency response curve) the pulse shape of a photon hit (Fig. 2.b) to show two different decay times.*

that if the two decay times and the ratio between the amplitudes are fixed at the same values for all photons, this simplified double-exponential model still replicates the complex pulse shapes sufficiently. We use it to fit every phase pulse to obtain the phase shift at the moment of absorption, which is proportional to the photon energy. We expect to be able to further improve performance by developing a fit that exploits the unique shapes of both the resonance frequency versus temperature dependence as well as the frequency response. We design our TKIDs with long signal decay times even though those reduce the maximum possible count rate – if a second photon hits on the decaying tail of a previous one, reconstructing its energy becomes difficult. However, long decay times also reduce phonon noise[25], help to average down HEMT and TLS noise[26] and allow more accurate fits to the pulses, increasing the accuracy of photon energy reconstruction. It is straightforward to engineer decay times to a desired maximum count rate by varying the thermal link between island and temperature bath.

The final energy resolution of one of our prototype TKIDs is demonstrated in Fig. 4, showing 4970 photons collected at a base temperature of 170 mK. We use an $Fe^{55}$ X-ray source and, fitting the spectrum with three normal distributions centered at the Mn K-α and K-β lines, we calculate an energy resolution of 75 eV at 5.9 keV. This is the first time for TKIDs to actually demonstrate energy resolution.

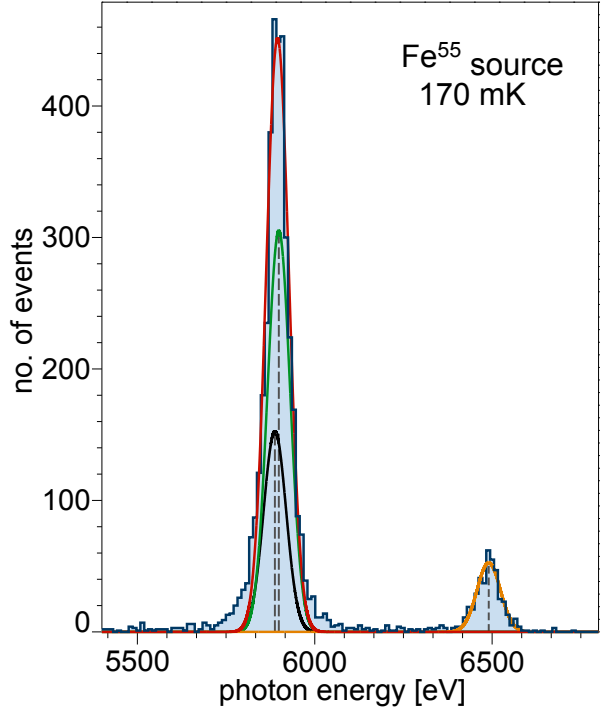

*Fig. 4: Distribution of the fitted energies of 4970 absorbed X-ray photons from an Fe$^{55}$ source, measured at 170 mK. The close Mn K-α doublet at 5.888 keV (black fit) and 5.899 keV (green) is not resolved, but the splitting between the combined 5.9 keV line (red) and Mn K-β at 6.49 keV (orange) is clear. From the line splitting and the average line width we calculate an energy resolution of 75 eV at 5.9 keV.*

The operating temperature $T_0$ can be used to customize a TKID for desired sensitivity and photon energy. As our TKIDs use superconducting Ta absorbers, their heat capacities are small but rise quickly with membrane temperature (C ~ $T^3$ as we operate well below the Ta $T_c$ of about 4.5 K). Furthermore, the membrane temperature rise we expect is large compared to our operating temperature, thus modest changes in $T_0$ have little effect on the maximum island temperature after photon absorption. Hence lowering $T_0$ results in bigger pulses (see Fig. 3.c) and increases sensitivity and signal to noise ratio. Higher $T_0$ decreases pulse heights but allows for measurement of higher photon energies. However, if $T_0$ is not sufficiently below the $T_c$ of TiN$_x$, $Q_i$ decreases and noise increases.

A 6 keV photon increases our TKID island temperature by about 200 – 300 mK, resulting in a pulse height of more than 150°. As the slope of the frequency response curve gets smaller above phase shifts of about 100° (see Fig. 3.a) this reduces our energy resolution. We reduce this saturation effect by working with a TiN$_x$ $T_c$ of 1.6 K and operating at 170 mK, but we can't eliminate it completely as the decline of $Q_i$ starts to dominate for higher $T_0$ and a higher TiN$_x$ $T_c$ reduces the island's heat capacity too much. This means our prototype TKIDs are still too sensitive, and we should be able to increase energy resolution by optimizing their thermal design. The straightforward way to reduce the sensitivity of a TKID is to increase its absorber's heat capacity, but first experiments with Au absorbers failed. As TKIDs are driven at resonance, the current density in the inductor is high and can induce substantial eddy currents in the nearby absorber. We think these eddy currents heat a non-superconducting absorber significantly and are the reason our initial Au absorber devices didn't work. Finding the optimal tradeoff between TiN$_x$ $T_c$, sensitivity, saturation, and noise is work in progress. Preliminary calculations show that a device with an optimized thermal design utilizing the steepest part of the frequency response curves of Fig. 3.a should be able to resolve less than 10 eV at 5.9 keV[19]. Further optimization of sensor fabrication and thermal design should thus bring TKIDs into the energy resolution range currently achieved by TES microcalorimeters, facilitating significant progress in both astronomical X-ray observations with next generation telescopes as well as high resolution X-ray spectroscopy at synchrotron light sources.

The authors would like to thank the NASA Space Technology Research Fellowship (NSTRF) program (grant NNX13AL70H) for funding graduate student P. S.. The work was funded by the NASA ROSES-APRA detectors program grants NNX13AH34G and NNX14AI79G. Devices were made at the UC Santa Barbara Nanofabrication Facility, a part of the NSF funded National Nanotechnology Infrastructure Network.

# Supplementary Material for:
# Highly Multiplexible Thermal Kinetic Inductance Detectors for X-Ray Imaging Spectroscopy

Gerhard Ulbricht[1,a)], Benjamin A. Mazin[1], Paul Szypryt[1], Alex B. Walter[1], Clint Bockstiegel[1], Bruce Bumble[2]

[1] Department of Physics, University of California, Santa Barbara, California 93106, USA
[2] NASA Jet Propulsion Laboratory, 4800 Oak Grove Drive, Pasadena, California 91125, USA

*TKID fabrication:*

Our TKIDs are fabricated in a 5 layer lithographic process. For the substrate, we use an undoped high resistivity 4" Si wafer coated with 20 nm $SiO_2$ and 250 nm $Si_3N_4$. The $Si_3N_4/SiO_2$ layer is patterned and etched at 100 W in a 0.4 mbar $CF_4 + O_2$ plasma. We preserve the $Si_3N_4$ underneath the inductor and absorber to later form the free-standing island but remove it underneath the capacitor as this significantly improves the resonator's noise performance[1]. As we need a high $Q_i$, we use a dedicated UHV sputter system at UCSB to deposit 200 nm $TiN_x$ by sputtering ultra-pure Ti in an $Ar + N_2$ (40 : 3) atmosphere at $1 \cdot 10^{-2}$ mbar and 500 W. This results in a Ti to N ratio that gives a $T_c$ of 1.6 K[2] and proved to be optimal for our current TKID design. Optimized sputter conditions are crucial for both $T_c$ control and high $Q_i$, and we are able to reach $Q_i$ values of over $3 \cdot 10^5$. The $TiN_x$ is structured by ICP etching in $BCl_3 + Cl_2$ at $2 \cdot 10^{-2}$ mbar. To allow maximum flexibility with film thickness and $T_c$, only the resonators themselves are made of $TiN_x$. We fabricate feed line and ground plane in the next step by lift-off of a 130 nm Nb film, sputtered at 120 W in $5 \cdot 10^{-3}$ mbar Ar. Subsequently we sputter 500 nm Ta (120 W, $5 \cdot 10^{-3}$ mbar Ar) and pattern it to form the absorbers, also by lift-off. Underneath the Ta we deposit about 15 nm of Nb as seed layer. In the last lithographic step, we structure photo resist to protect most of the sample's surface during $XeF_2$ etching. The wafer is then diced and single chips are etched in $XeF_2$ (five 30 s cycles at 4 mbar) to release the free-standing islands – the $SiO_2$ underneath the $Si_3N_4$ protects it from the $XeF_2$. The $Si_3N_4$ bridges would be sufficiently rigid to remove the protective resist in solvents, but as the $XeF_2$ seems to harden it substantially, we burn the resist off in an oxygen plasma. Particles from dicing are no problem as the $XeF_2$ removes them. Our prototype chips have 38 TKIDs each with resonance frequencies between 2.5 GHz and 5.5 GHz, and in most cases all of them can be addressed.

*Figure 2.a explained in more detail:*

When we measure a frequency sweep like the blue line in Fig. 2.a, we get a combination of a small loop in the IQ plane for the resonator and a bigger loop caused (among other things) by the feed line. However, during the actual measurement, the drive frequency is fixed and any feed line components in the data are constant. Therefore, in order to extract the frequency response curves of Fig. 3.a+b, we removed these feed line components by fitting the big loop and subtracting it. The purpose of Fig. 2.a is to illustrate how we extract the phase shift a photon hit generates from measured data. For the frequency sweep values (blue line) we show a sweep measured with 7 dB less power than the pulse data (red dots), as high power frequency sweeps often show confusing artifacts where the resonator loop merges with the feed line loop. Therefore we had to scale the frequency sweep data by 7 dB = 5.012 and to shift it in the IQ plane. We made sure that this shift reproduced the typical behavior of observed pulse data with respect to frequency sweeps. However it is not sufficiently accurate to draw conclusions from Fig. 2.a about the pulse shape if instead of phase the TKID's resonance depth (dissipation) would be plotted. A much more careful analysis is necessary to extract a TKID's dissipation behavior.

*Estimating the optimized TKID energy resolution:*

To estimate the energy resolution of an optimized TKID we compare the highest slope of the frequency response curve in Fig. 3.a with its slope 2.7 MHz below the drive frequency ($\Delta f_r = -2.7$ MHz is the extrapolated value for a 350 mK island temperature from Fig. 3.c). This would result in a possible energy resolution enhancement by a factor of 60. There are several uncertainties in this extrapolation and increasing the TKID sensitivity also increases some (but not all) noise sources. Therefore we estimate that an increase in energy resolution of at least one order of magnitude should be attainable. Fundamental limits, like phonon noise or generation-recombination noise, are (for T below 200 mK) lower than 2 eV at 6 keV.